\documentclass[12pt,onecolumn,aps,pra,floatfix,superscriptaddress,preprint, showpacs,groupedaddress]{revtex4-1}

\usepackage{graphicx}
\usepackage{graphics}
\usepackage{amssymb}
\usepackage{amsmath}
\usepackage{epsfig}
\usepackage{latexsym}
\usepackage{color}
\usepackage{rotating}
\usepackage{subfigure}
\usepackage{float}
\usepackage[breaklinks=true]{hyperref}
\usepackage{siunitx}

\usepackage{setspace}

\hypersetup{
  colorlinks   = true, 
  urlcolor     = blue, 
  linkcolor    = blue, 
  citecolor   =  red 
}

\usepackage{soul} 

\begin{document}

\title{Ultrafast viscosity measurement with ballistic optical tweezers}

\author{Lars S. Madsen$^{1,*}$, Muhammad Waleed$^{1,*}$, Catxere A. Casacio$^1$, Alexander B. Stilgoe$^2$, Michael A. Taylor$^3$,
  and Warwick P. Bowen$^{1,}$\footnote{w.bowen@uq.edu.au}}
\affiliation{$^1$ARC Centre of Excellence for Engineered Quantum Systems, University of Queensland, St Lucia, QLD 4072, Australia}
\affiliation{$^2$School of Mathematics and Physics, University of Queensland, St Lucia, QLD 4072, Australia}
\affiliation{$^3$Australian Institute for Bioengineering and Nanotechnology, University of Queensland, St Lucia, QLD 4072, Australia}
\affiliation{$^*$ {\rm These authors contributed equally.}}

\begin{abstract}
Viscosity is an important property of out-of-equilibrium systems such as active biological materials and driven non-Newtonian fluids, and for fields ranging from biomaterials to geology, energy technologies and medicine. However, noninvasive viscosity measurements typically require integration times of seconds. Here we demonstrate a four orders-of-magnitude improvement in speed, down to twenty microseconds, with uncertainty dominated by fundamental thermal noise for the first time. We achieve this using the instantaneous velocity of a trapped particle in an optical tweezer. To resolve the instantaneous velocity we develop a structured-light detection system that allows particle tracking with megahertz bandwidths. Our results translate viscosity from a static averaged property, to one that may be dynamically tracked on the timescales of active dynamics. This opens a pathway to new discoveries in out-of-equilibrium systems, from the fast dynamics of phase transitions, to energy dissipation in motor molecule stepping, to violations of fluctuation laws of equilibrium thermodynamics. 
\end{abstract}

\maketitle


Optical tweezers provide a unique tool to study the basic mechanics of life. By tracking, trapping and manipulating micrometre-sized particles in solution~\cite{ashkin1986observation}, they have been used to measure the stepping of individual motor molecules~\cite{svoboda1993direct,carney2020direct}, the strength of single DNA strands~\cite{wang1997stretching}, and the active material properties of living cells~\cite{nishizawa2017feedback} among many other applications~\cite{killian2018optical,tassieri2019microrheology}. However, even at the shortest timescales available to these experiments, the thermally driven motion of the particle is well described by random Brownian diffusion. As such, the instantaneous velocity of the particle is inaccessible.
Access to the particles instantaneous velocity allows
direct microscale measurements of kinetic energy and energy dissipation, and therefore also of local viscosity~~\cite{grimm2012high}.
Importantly, the velocity of a trapped particle reaches equilibrium with the  liquid around it over exceedingly short timescales,  typically orders-of-magnitude shorter than is the case for position~\cite{kheifets2014observation}.
This offers the prospect to observe fast particle-liquid interactions and to probe the properties of the liquid 
at higher rates than is currently possible.

Improved measurement speeds
are particularly needed
for out-of-equilibrium systems, such as active biological materials driven by the action of motor molecules~\cite{gnesotto2018broken, dogterom2019actin, nishizawa2017feedback}, and non-Newtonian fluids under dynamic loads~\cite{rathee2017localized, waitukaitis2012impact},  which commonly exhibit rich dynamics with high spatial and temporal resolution~\cite{han2016high, saint2018uncovering, tassieri2019microrheology}. They are also needed to improve understanding of fast single-molecule dynamics and enzymatic activity~\cite{capitanio2012ultrafast}.
Indeed, the disparity between the seconds-to-minutes timescales of typical measurements 
with optical tweezers 
and the sub-second timescales of active biological processes, have led authors to conclude that  these measurements are ``not an option" for living cells~\cite{tassieri2015linear}. 
Direct measurements of energy exchange on shorter timescales, and particularly of local viscosity, can be expected to provide new insights into the thermodynamics of out-of-equilibrium systems, ranging from the efficiency of biological machines~\cite{ariga2018nonequilibrium}, to violations of fluctuation theorems~\cite{battle2016broken, gnesotto2018broken}, soft matter phase transitions~\cite{oyama2019glassy,guzman2008situ} and chaotic active behaviours~\cite{grob2016rheological}.

Here we experimentally validate that access to the instantaneous velocity allows the energy exchange between a trapped particle and fluid to be probed at a high rate. This allows us to track the viscosity of the fluid with temporal resolution down to 20~$\mu$s, more than four orders-of-magnitude faster than has been demonstrated previously~\cite{pralle1998local}, and with uncertainty dominated by the fundamental thermal noise floor for the first time. We resolve the instantaneous velocity 
 by developing  a new structured-light detection approach for optical tweezers, which filters the bright background of the optical trap while amplifying the signals generated by particle motion. This method allows higher speed particle tracking than has been possible previously~\cite{kheifets2014observation}.  It also removes the need for custom high-power detectors~\cite{chavez2008development}, making it widely accessible. 
Our results change the paradigm of viscosity measurements from static background measurement to a dynamic variable that can track the local changes in the particles surroundings, providing a new tool to answer both fundamental and applied questions in the  dynamics of out-of-equilibrium systems.

\section{Results}

\subsection{Velocity thermalisation allows fast viscosity measurements}

\begin{figure}
\begin{center}
\includegraphics[width=1\textwidth,keepaspectratio]{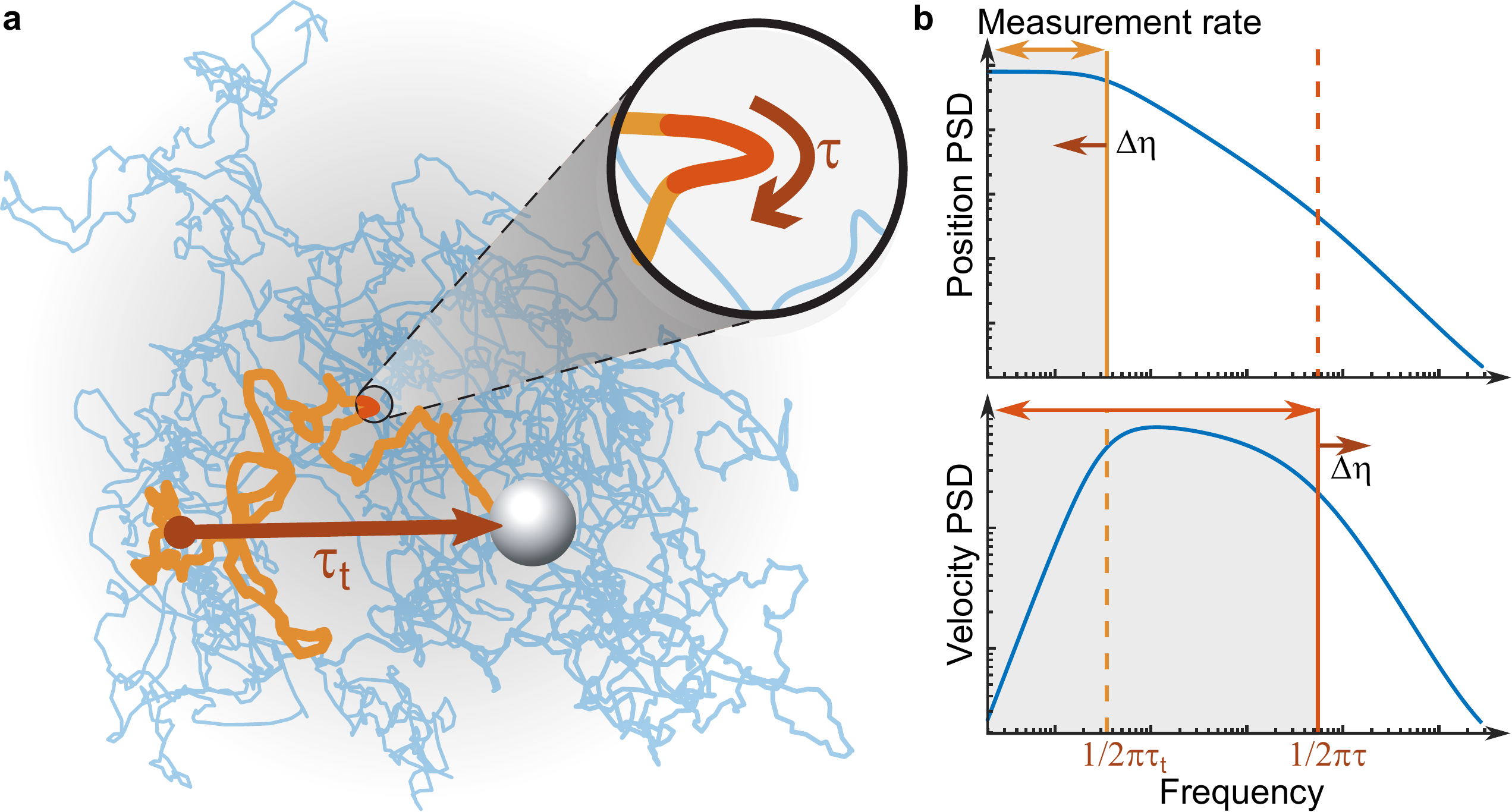}
\caption{
\footnotesize  \textbf{Fast velocity thermalisation increases the speed of viscosity measurements.}
\textbf{a,} A bead trapped in an optical tweezers performs thermally-driven motion. The position of the particle thermalises over the position relaxation time $\tau_t$, which quantifies the time taken on-average for the particle to traverse the trap and explore its potential energy landscape. The particle velocity thermalises on the faster characteristic timescale $\tau$, allowing kinetic energy to be explored at a much higher rate (inset). 
\textbf{b,} Position and velocity power spectral densities (PSDs). Tracking the corner frequency introduced by position relaxation (yellow line) allows viscosity measurements, with a maximum detection rate given by the corner frequency itself (grey shading, {\it top}).  Tracking the corner frequency due to velocity relaxation (orange line) allows a maximum detection rate that is several orders of magnitude higher (grey shading, {\it bottom}).
}
\label{fig1}
\end{center}
\end{figure}

The viscosity $\eta$ of a fluid can be obtained statistically from the trajectory of a particle in an optical tweezer, as illustrated in Fig.~\ref{fig1}{\bf a}.  Passive methods generally estimate the trap frequency of the optical tweezer $f_t$  from the corner it introduces in the position power spectral density of the particles thermally driven motion~\cite{pralle1998local} (see Fig.~\ref{fig1}{\bf b} ({\it top})). Hydrodynamic theory connects the trap frequency to viscosity, e.g. through the relation $\eta = \kappa/(12 \pi^2 r f_t)$ for the usual case of a spherical particle of radius $r$ and a trap of stiffness $\kappa$~\cite{grimm2012high}. However, sequential estimates of $\eta$ are only statistically independent 
   if the time between them is longer than the position relaxation time 
 $\tau_t = 1/2 \pi f_t$. $\tau_t$ is typically in the range of 0.1~to~10~ms and quantifies the timescale for the particle to diffuse from one side of the trap to the other. This  severely constrains the speed of the measurement, with seconds-to-minutes long integration times required to average uncertainty down to acceptable levels~\cite{tassieri2015linear,  tolic2006calibration, pralle1998local, bishop2004optical}.
Active methods use an external force to drive the particle motion~\cite{tolic2006calibration, guzman2008situ, le2010simultaneous}, but face a similar problem --- the position decorrelation time determines the maximum speed with which the particle position can respond to the applied force, and therefore the maximum measurement rate.

In the ballistic regime, for which the measurement is fast compared to the particles momentum relaxation time $\tau_i$, the viscosity can be obtained more directly through its connection to  kinetic energy dissipation. Neglecting hydrodynamic memory effects, collisions with molecules in the fluid exponentially damp the velocity of the particle over the momentum relaxation time. This momentum relaxation typically occurs at a much faster rate than the position relaxation, as illustrated in Fig.~\ref{fig1}{\bf a} --- that is to say, the velocity of the particle equilibrates with the fluid environment much more quickly than does its position. This is predicted to have significant consequences for out-of-equilibrium systems. For example, it causes the effective temperatures of the position and velocity of a hot Brownian particle to diverge~\cite{joly2011effective, geiss2019brownian}. Here, we use it to reduce the integration time required for precise viscosity measurements, and therefore to increase the speed of the measurement. 

For a spherical particle, the fluid viscosity is related to the momentum relaxation time via $\eta = m^*/6 \pi r \tau_i$~\cite{kheifets2014observation}, where $m^*=m+m_f/2$ is the effective mass of the particle, increased due to the inertia of the fluid envelope that surrounds it, $m$ is its actual mass and $m_f$ is the mass of fluid it displaces as it moves. This illustrates a second advantage of exploiting the ballistic motion to measure viscosity~\cite{grimm2012high} --- that the conversion from relaxation time to viscosity is independent of all optical tweezers trapping parameters. It only depends on the shape of the particle and the densities of the particle and fluid, and is therefore immune to many possible sources of calibration fluctuation such as those associated with focussing through a spatially structured biological media or with trap intensity drift.

The simple physics discussed above is complicated by hydrodynamic memory effects due to vorticity generated as the particle moves through the fluid. This modifies both the characteristic time $\tau$ for dissipation and  the functional form of the decay~\cite{kheifets2014observation, alder1970decay}. At short times ($t \ll \tau$), the decay is well approximated by the exponential of a square-root ($e^{-\sqrt{t/\tau}}$) with
$\tau =  4\tau_i m^* \!  /9m_f$. At longer times, 
the fluid vorticity produced by the motion of the particle introduces a correlated driving force leading to memory effects and velocity correlations~\cite{franosch2011resonances, huang2011direct}.
Nevertheless, the qualitative conclusions remain the same, that the fast equilibration of the particle velocity with the fluid allows fast viscosity measurements that are insensitive to fluctuations in the optical trap.

The hydrodynamic memory of the driving force can be deconvolved by moving into the frequency domain. For this reason, as well as for technical reasons, it is convenient to analyse the particle motion in the frequency domain (see Ref.~\cite{berg2004power} and Supplementary Information Section 1). 
Similarly to position relaxation, the momentum relaxation introduces a corner in the velocity power spectral density at a frequency around $1/2 \pi \tau$, as shown in Fig.~\ref{fig1}{\bf b} ({\it bottom}). This corner becomes resolvable once the optical tweezer has sufficient precision to reach the ballistic regime. The corner frequency depends sensitively on the viscosity of the surrounding fluid (through $\tau$).  Through this dependence, a fit to the 
power spectral density provides a means to determine the viscosity.

\subsection{Structured-light detection in optical tweezers}

   \begin{figure}[htb!]
\begin{center}
\includegraphics[width=0.5\textwidth,keepaspectratio]{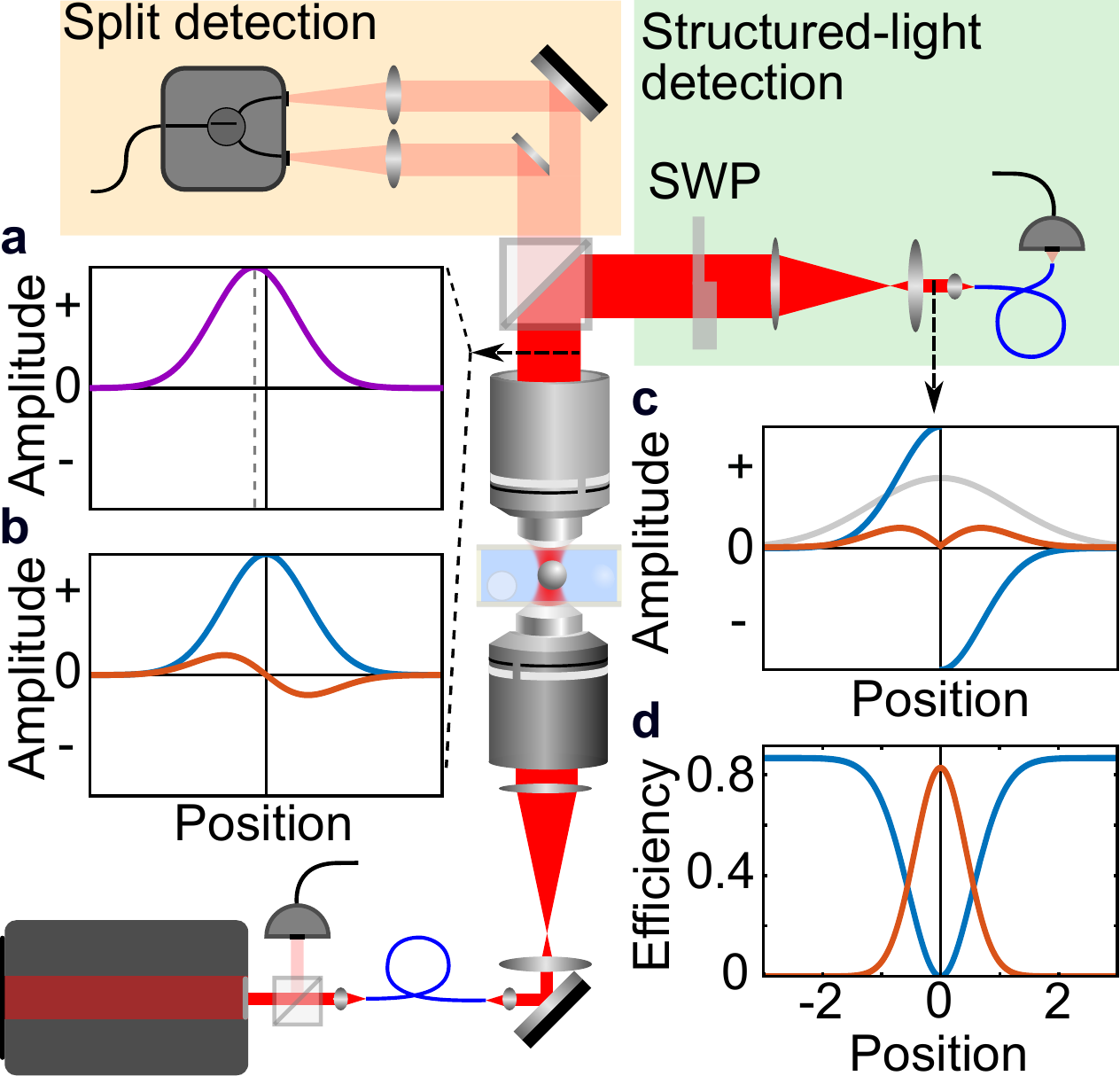}
\end{center}
\caption{\footnotesize  \textbf{Optical tweezers with structured-light detection.}
The optical tweezer incorporates both structured-light detection (green shaded box) and conventional split-detection (orange shaded box). The former is used to reach the ballistic regime, and the latter to characterise the low frequency Brownian motion of the particle.  SWP: split waveplate.
\textbf{a-c,} Structured-light detection concept. \textbf{a,} Transverse field amplitude in the image plane after the optical tweezer, shifted due to the particle displacement (dashed line). \textbf{b,} The displaced field amplitude can be decomposed into a symmetric mode (blue) and an information-carrying antisymmetric mode (red). 
\textbf{c,} The split waveplate flips the phase of one half of the field, switching the symmetry. 
 \textbf{d,} Coupling efficiency of symmetric and antisymmetric modes into single-mode fibre as a function of transverse position of split waveplate. Approaching the centred position the symmetric component (blue) is suppressed and the ideal coupling efficiency of the displacement signal (red) reaches $83\%$.
 }
\label{fig2}
\end{figure}

To reach the ballistic regime, the thermal motion of the tracked particle must be resolved in a time shorter than the momentum relaxation time. This is highly nontrivial for a microparticle in a typical liquid, both because the relaxation occurs over sub-microsecond timescales, and because the interaction of light with the particle is weak. The weakness of the interaction can be understood by considering how the pattern of light scattered from a particle changes as it moves away from the axis of a focussed beam. In thermal equilibrium the particle travels an average distance of around $\sqrt{k_B T/m^*} \, \tau \sim 100$~pm within the momentum relaxation time, where $k_B$ is Boltzmann's constant. Since this is far smaller than optical wavelengths, even in the ideal case 
of diffraction-limited focussing
 the scattering pattern is essentially identical to that of an on-axis particle. Indeed, only around 
one photon in a billion carries information about the particles position (see Supplementary Information Section 2). In an optical tweezer, this necessitates the use of a  high power trapping field, together with efficient techniques to collect information from  the scattered photons that introduce minimal noise from the trapping field, electronics and vibrations.

The pioneering experiments which first reached the ballistic regime, and remain the state-of-the-art, used custom high-power split-detectors that were capable of detecting
around 100~mW of trapping light with shot-noise limited performance~\cite{kheifets2014observation}. In this approach, the microparticle motion is observed due to the interference of the information-carrying component of the scattered light with the far brighter transmitted trapping beam, which acts to displace the trapping beam on the split-detector.

Here, we develop an alternative approach which avoids the need to detect the trapping field, and therefore can both be combined with an arbitrarily strong trap, and implemented with commercial off-the-shelf detectors.
Our approach uses structured-light detection to spatially filter the information-containing component of the scattered field from the trapping field, as shown in Fig.~\ref{fig2} (see Methods for details). The key concept is that the transverse amplitude profile of the displaced trapping beam can be decomposed into a component that is symmetric on reflection about the centre of the trap, predominantly from the unscattered trap field, and an anti-symmetric component that carries all the information about the position of the particle. Insets {\bf a}\&{\bf b} in Fig.~\ref{fig2}  illustrate this concept. The power contained in the anti-symmetric component is zero when the particle is at the centre of the trap, and increases as it moves to either side, with a $\pi$ phase shift in its amplitude profile distinguishing whether the particle is on the left or right side. We take advantage of the difference in symmetry to filter out the trapping field.    Compared with split-detection, our structured-light detection approach not only resolves the problem of power handling, but also allows the mode-shape of the detected field to be optimised improving the detection efficiency (see Methods).

To implement the spatial filter we insert a custom-designed split half-waveplate into the back-focal plane of the microscope~\cite{treps2003quantum, taylor2013biological}. The waveplate is cut into quarters and reassembled to introduce a $\pi$ phase shift to the light on one side of the trap axis, while leaving light on the other side unchanged. This reverses the symmetry -- the trapping field becomes anti-symmetric, while the information-containing field becomes symmetric (see Fig.~\ref{fig2}{\bf c}). 
The combined field is then aligned  onto a single-mode optical fibre. Since the guided mode of the fibre is Gaussian (and therefore symmetric), when the alignment is perfect the information-containing field is maximally coupled and the trap field is fully rejected (see Fig.~\ref{fig2}{\bf d}). In practise, we find that the suppression of the trap field can be higher than $10^3$. This allows hundreds of milliwatt trapping fields to be combined with off-the-shelf broadband detectors that saturate at around a milliwatt of power (see Supplementary Information Section 4.1).  Inspired by near-dark-fringe detection in gravitational wave interferometers~\cite{meers1988recycling}, where gravitational wave signals are boosted above the electronic noise by operating the
interferometer near -- but not at -- the dark-fringe, we deliberately introduce a small misalignment. This allows a small fraction of the trapping field to enter the fibre, acting as a local oscillator that amplifies the position signal from the scattered field above the electronic noise floor of the detector.

\subsection{Reaching the ballistic regime}

\begin{figure}[htb]
\begin{center}
\includegraphics[width=1\textwidth,keepaspectratio]{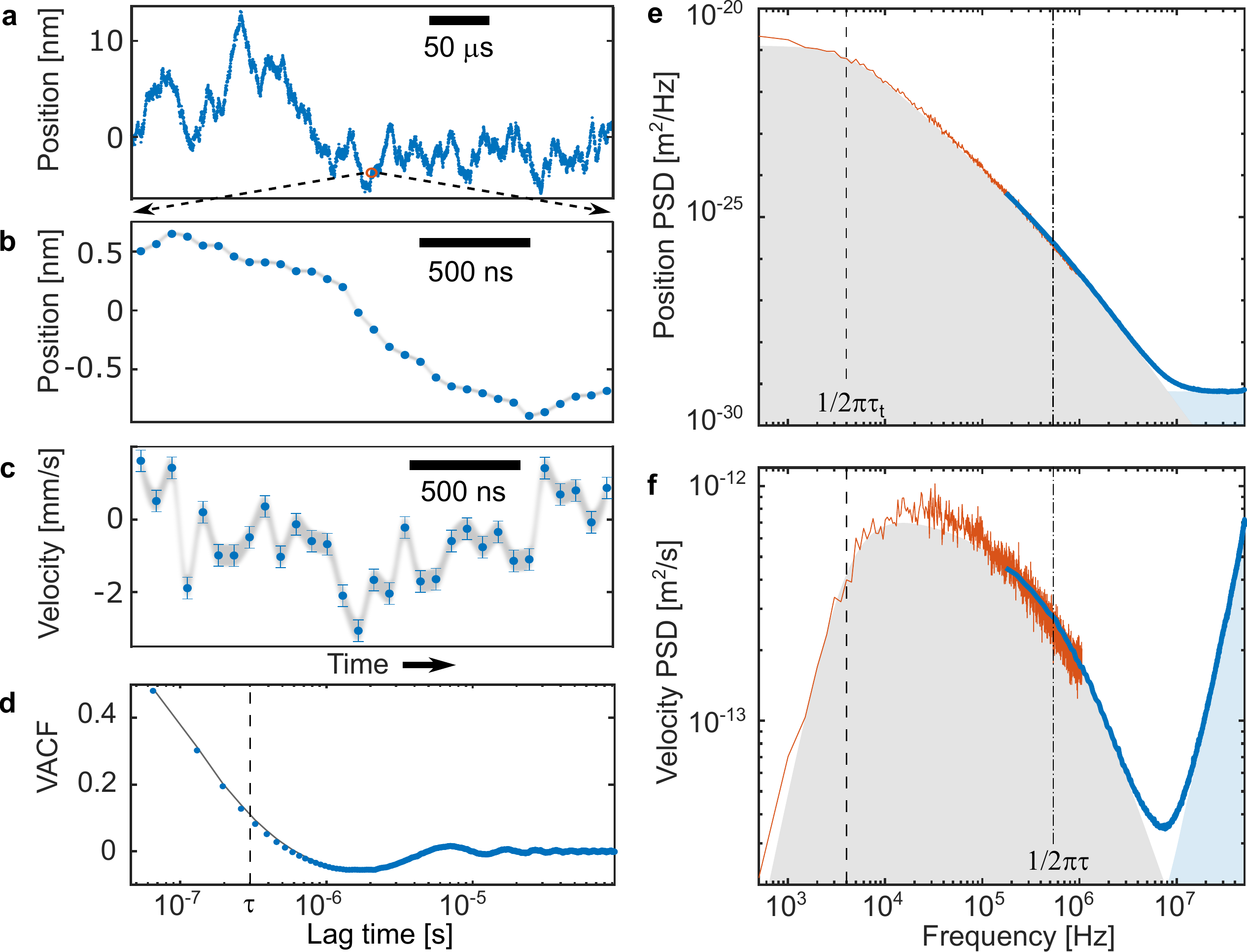}
\end{center}
\caption{\footnotesize  \textbf{Microparticle tracking in the ballistic regime.}  {\bf a,} Position trajectory of a 1.18~$\mu$m diameter silica microsphere in water taken with conventional split-detection over a time spanning $10 \tau_t$. {\bf b,} High resolution trajectory taken simultaneously with structured-light detection, shown here over a time of $10 \tau_i$. 
{\bf c,} Particle velocity calculated from the data in {\bf b}. Error bars: one-sigma uncertainty due to laser noise obtained from simulations.
{\bf d,} Velocity autocorrelation function calculated from data in {\bf c,} (blue points) compared to theory (grey line). Note: the oscillations are an artefact arising from highpass filtering.  
{\bf e} \& {\bf f,} Position and velocity power spectral densities, calculated as described in Section 1 of the Supplementary Information. The low frequency components (red traces) were obtained with split-detection, and the high frequency components (blue traces) were obtained with structured-light detection. Grey shading: theoretically predicted power spectra from thermal motion alone. Blue shading: noise floor of structured-light detection. Dashed line: $1/2 \pi \tau_t$. Dot-dashed line: $1/2 \pi \tau$. 
}
\label{fig3}
\end{figure}

To test the performance of our structured-light detection scheme we first track the dynamics of 1.18~$\mu$m diameter silica microspheres in water using 200~mW of trapping power at the sample with 100 mW reaching the structured detector. The characteristic time for momentum relaxation can be determined using the particle mass $m_{\rm p} = 4 \pi \rho r^3/3 = 1.7$~pg, the mass of the displaced fluid envelope $m_{\rm f} = 2 \pi \rho_f r^3/3 = 0.43$~pg, and the known viscosity of water of $\eta = 0.94$~mPa$\cdot$s at 296~K, where $\rho$ and $\rho_f$ are the density of silica and the fluid respectively.
For this example choice of particle and fluid, we find $\tau = 300$~ns which, compared to $\tau_i$, is only increased by around 10\% due to hydrodynamic memory effects. By comparison, using the trap-stiffness $\kappa = 230~\mu$N/m obtained from a fit to the experimental power spectrum, we find a more than  two order-of-magnitude longer position relaxation time of $\tau_t = 45~\mu$s. This suggests that significant gains are possible within the ballistic regime.

An example of the observed position trajectory of the particle as it moves over time is shown in Fig.~\ref{fig3}{\bf a}. Ordinary Brownian motion is seen over long timescales. Over shorter timescales, however, the motion displays smooth trajectories characteristic of ballistic motion, as shown in Fig.~\ref{fig3}{\bf b}. This allows the instantaneous particle velocity to be extracted (Fig.~\ref{fig3}{\bf c}, see Section 1 of the Supplementary Information for details). The velocity autocorrelation function calculated from the particle velocity shows significant autocorrelation due to inertia at times beneath the momentum relaxation time, as quantified in Fig.~\ref{fig3}{\bf d}. We thus conclude that our structured-light detection scheme provides access to the ballistic regime.

The position and velocity power spectral densities are calculated from the Fourier transform of the particle position trajectory (blue traces in Fig.~\ref{fig3}{\bf e}\&{\bf f}). Achieving the 90~dB  of dynamic range required to resolve the motion at frequencies both below the trap frequency and above the ballistic corner frequency is not technically feasible with our apparatus. We therefore use a high-pass filter to remove low frequencies in data acquisition (see Supplementary Information Section 4.1), using standard split-detection to fill in this part of the spectrum (orange traces). As can be seen in Fig.~\ref{fig3}{\bf e}\&{\bf f}, excellent agreement between data and theory is achieved across the full measured spectrum. Importantly, our structured-light detection scheme allows the thermal motion of the particle to be resolved above the optical shot-noise (blue shading) over a bandwidth up to 7.4~MHz, an order of magnitude higher than the momentum relaxation corner frequency and therefore deep within the ballistic regime.

\subsection{Absolute and ultrafast viscosity measurements}

\begin{figure}
\begin{center}
\includegraphics[width=1\textwidth,keepaspectratio]{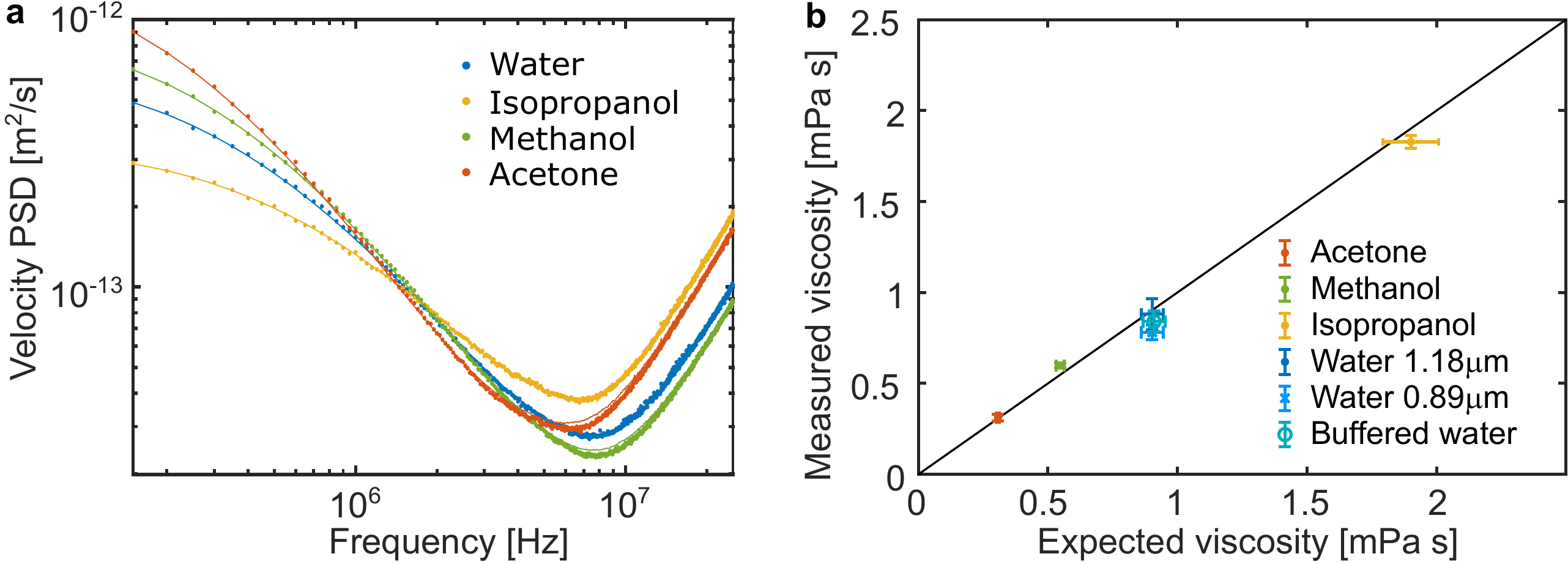}
\end{center}
\caption{\footnotesize \textbf{Absolute viscosity estimation.}
\textbf{a,} Examples of the velocity power spectral densities used to estimate the viscosity of acetone, methanol, water and isopropanol. Points: data taken with 1.18~$\mu$m microspheres over a 0.2~s acquisition time. Lines: corresponding fits.  
 \textbf{b,} Comparison of viscosity obtained from the fit to the expected value for each liquid. Data is also shown for a buffered water solution, and using a smaller $0.89~\mu$m diameter particle in water. 
 To quantify the statistical uncertainty each measurement is repeated for multiple particles and multiple times for each particle.
Repeated measurements on a single particle reveal
 fluctuations that are attributed to debris in the solutions, while discrepancies between 
particles are attributed to variations in size and shape. Overall, these two sources of uncertainty have similar magnitude (see Section 4.4 of the Supplementary information). The vertical error bars represent their combined standard deviation. The horizontal error bars correspond to $\pm2$~K in temperature.
 Black line: guide to the eye. 
 }
\label{fig4}
\end{figure}

To estimate absolute viscosity from the velocity power spectral density, we perform a least-squares fit to the theoretically expected velocity power spectrum, using the viscosity and an overall scaling factor as fitting parameters (see Methods and Section 4.3 of the Supplementary Information for details). We find that to achieve accurate estimates, it is necessary to include not only hydrodynamic memory effects, but also the hydrodynamic coupling to the sample chamber wall (known as Faxen's correction~\cite{berg2004power}),
 the experimentally measured noise floor, and the transfer function of the detector (Supplementary Information Section 4.2).
Fig.~\ref{fig4}{\bf a} shows examples of fits obtained in acetone, methanol, water and isopropanol (Section 3.3 of the Supplementary Information), the tabulated viscosities of which vary from 0.3 to 1.9~mPa$\cdot$s~\cite{viswanath2007viscosity}. 
Fig.~\ref{fig4}{\bf b} compares the viscosities obtained 
 from the fits
  to the tabulated values, including corrections for laser heating (see \cite{peterman2003laser} and Methods).
The level of agreement in all cases, both for the fits and for the final viscosity estimates, demonstrates that our method provides accurate estimates over a wide range of viscosities.

\begin{figure}
\begin{center}
\includegraphics[width=0.4\textwidth,keepaspectratio]{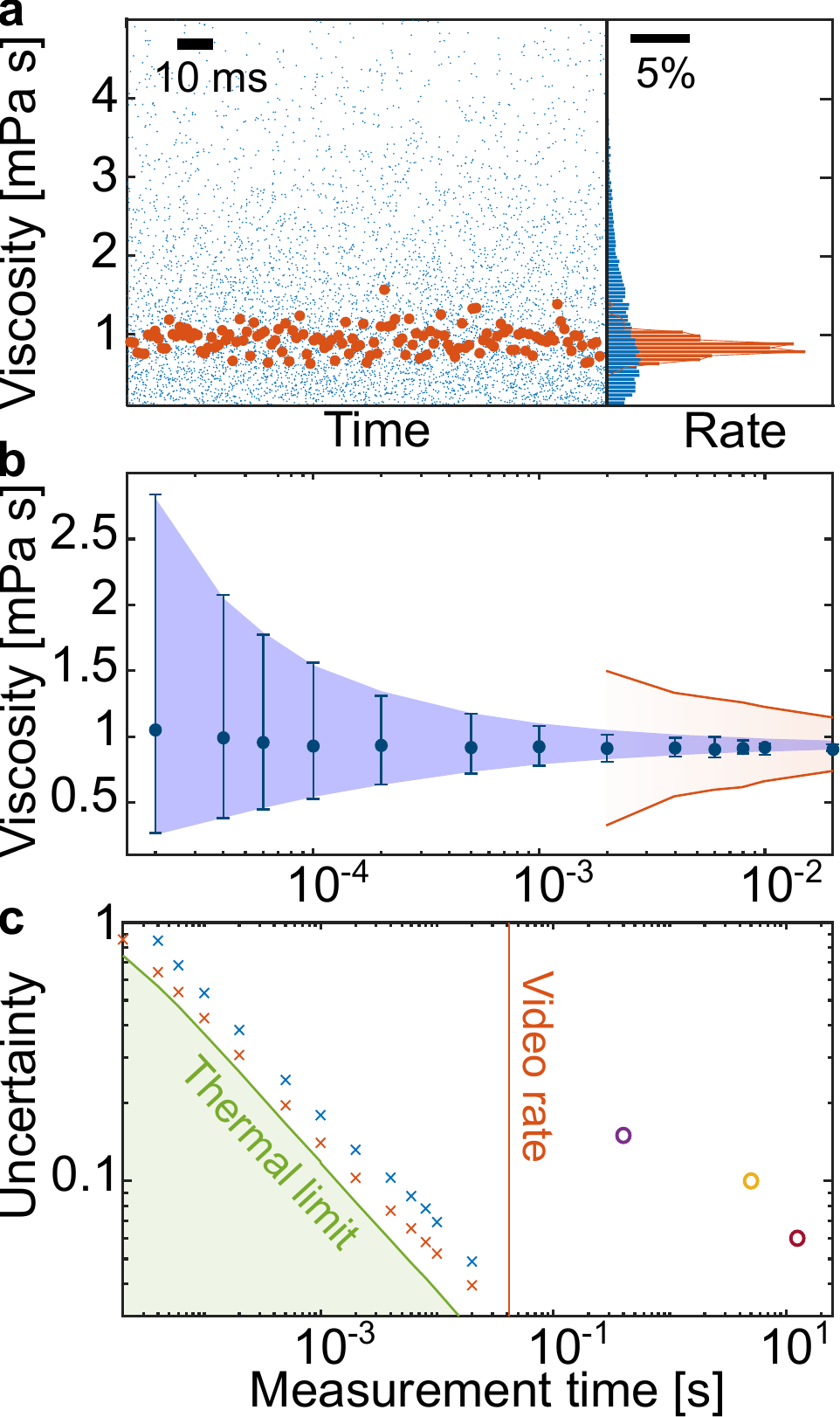}
\end{center}
\caption{\footnotesize  \textbf{Orders-of-magnitude faster viscosity measurement in the ballistic regime.}
\textbf{a,} {\it Left panel}: dynamic viscosity tracking with 1~ms (red) and 20~$\mu$s (blue) temporal resolution.
{\it Right panel}: probability distributions. \textbf{b,} Median and uncertainty of viscosity estimate as a function of temporal resolution. 
Error bars: 0.16 and 0.84 quantiles (corresponding to one sigma for a Gaussian distribution). 
Blue shading: quantiles from a simulation with identical parameters. 
Red lines: simulated quantiles when tracking the viscosity using the trap stiffness.  \textbf{c,} Standard deviation as a function of temporal resolution compared to the thermal noise limit and earlier studies.
Blue (red) crosses: observed uncertainty of measurements in buffered water solution using 40~mW (200~mW) of power on the sample. Green line: thermal noise-limited uncertainty obtained from simulated data. Red line: typical video rate of 24~Hz.  Circles: results from earlier passive (purple, \cite{pralle1998local}) and active (red, \cite{tolic2006calibration}; orange, \cite{bishop2004optical}) methods of viscosity estimation.
 }
\label{fig5}
\end{figure}

To test the speed and accuracy of the viscosity tracking, we acquire velocity data for a 1.18~$\mu$m silica microparticle in water using 40~mW of trap power at the sample. The data is divided into a series of short time segments, defining the temporal resolution of the measurement. We then estimate the viscosity separately for each time segment, 
using the full 0.2~s acquisition to calibrate the noise floor (see Supplementary Information Section 4.3 for more details). This provides high resolution traces of the viscosity as a function of time. 
Examples with temporal resolutions of 20~$\mu$s (blue data) and 1~ms (red data) are shown along with their probability distributions in Fig.~\ref{fig5}{\bf a}. Both measurements give an average viscosity consistent with the expectation of $\eta = 0.94$~mPa$\cdot$s for water at 296~K. The high resolution  20~$\mu$s  data has a large uncertainty and a skewed probability distribution, with estimated viscosity of $\eta = 1.1^{+1.7}_{-0.8}$~mPa$\cdot$s, where the uncertainty bounds define the 0.16 and 0.84 quantiles
(equivalent to one sigma for a Gaussian distribution). By comparison the additional averaging provided by using a 1~ms temporal resolution results in a probability distribution that is well approximated as Gaussian, with mean of $\eta=0.92$~mPa$\cdot$s and one-sigma uncertainty reduced to around 18$\%$.

Figure~\ref{fig5}{\bf b} shows the estimated viscosity and uncertainty as a function of temporal resolution, for resolutions ranging from 20~$\mu$s to 20~ms. As can be seen, the measurement provides median viscosities consistent with the expectation for water across this full range, with a monotonic reduction in uncertainty as the length of averaging increases. To confirm that the protocol performs as would be expected for an ideal experiment, we compare these results to simulated data for which the only sources of noise are the particles thermal noise and optical shot-noise (see Supplementary Information Section 1), shown with the solid blue band.
 The agreement indicates that the experiments are limited only by these fundamental noise sources, with any additional sources of error being insignificant over the full 0.2~s acquisition.
 
The simulations also allow an accuracy comparison against the usual approach for passive viscosity measurement, where a fit is used to extract the trap frequency~\cite{grimm2012high}. The simulated accuracy in this case as shown with the red lines in Fig.~\ref{fig5}{\bf b}. This shows that, compared to an ideal trap-based fit, the ability to measure within the ballistic motion regime allows a factor of five reduction in uncertainty at fixed temporal resolution, or equivalently twenty five times faster measurements at fixed uncertainty.
 
The thermal noise floor represents a fundamental limit to the performance of any viscosity estimation procedure, while shot-noise can be reduced by improving the optical apparatus. We determine the  thermal-noise-limited accuracy of the viscosity estimate for our procedure by running simulations that include no other noise sources. As shown in Fig.~\ref{fig5}{\bf c}, the accuracy, in this case, scales as the inverse-square-root of the temporal resolution, as expected from the central limit theorem. 
 To assess how closely our experiments approach this thermal noise limit, we determine the experimental uncertainty as a function of temporal resolution for two trapping powers, 40~mW and 200~mW at the sample, respectively.  As can be seen in Fig.~\ref{fig5}{\bf c}, in both cases the accuracy exhibits similar scaling to the thermal noise, and is only marginally degraded by shot-noise throughout the full range of temporal resolution studied. Indeed, the uncertainty is dominated by thermal noise and, once the ballistic regime is reached, is only marginally improved by increasing the trapping power. To take a specific example, for a 1~ms  time resolution, the thermal noise limited accuracy in the viscosity estimate is 12\%. With 200~mW of trap power shot-noise degrades the accuracy slightly to $14\%$, while reducing the power by a factor of five to 40~mW only degrades the accuracy to 18$\%$.

\section{Discussion}

As is expected due to the rapid thermalisation of velocity, our technique allows significantly faster viscosity measurements than other methods. For comparison, three indicative state-of-the-art results are included in Fig.~\ref{fig5}{\bf c}, two based on monitoring viscosity-induced changes in the trap frequency~\cite{tolic2006calibration, pralle1998local} and the other based on tracking its effect on the particles angular momentum~\cite{bishop2004optical}. To our knowledge, the fastest previously reported viscosity measurement in optical tweezers had a speed of 0.4~s~\cite{pralle1998local} (purple point in Fig.~\ref{fig5}{\bf c}). Methods used to determine the particle radius, from which the viscosity could also be estimated, have reported similar temporal resolution down to 0.35 s~\cite{grimm2012high, hammond2017direct, kheifets2014observation}.   By comparison, our technique achieves the same accuracy with a four hundred times shorter measurement duration, and importantly allows continuous viscosity measurements at more than four order-of-magnitude faster rates, albeit with increased uncertainty at these short times. In this way, our technique transforms viscosity from an averaged equilibrium property, into a parameter that can be monitored locally as it changes in real-time in out-of-equilibrium matter.

We expect the capability for real-time viscosity measurements developed here to provide new insights into a wide range of out-of-equilbrium systems. Deep questions remain around the connection between microscopic and macroscopic behaviour~\cite{gnesotto2018broken, comtet2017pairwise}, about the existence and applicability of fundamental thermodynamic laws~\cite{geiss2019brownian}, and about the mechanisms for observed behaviours~\cite{saint2018uncovering, brown2014shear}.  The speed of existing measurements is a key barrier to answering these questions. For instance, 
the complex spatiotemporal dynamics of active biological materials such as the cellular cytoplasm generally have timescales faster than are accessible with current viscosity measurements~\cite{arbore2019probing, koenderink2006high}. The averaging of these dynamics makes it highly non-trivial to draw unambiguous conclusions about the out-of-equilibrium characteristics of these systems~\cite{tassieri2015linear, arbore2019probing}. Similarly, non-Newtonian fluids exhibit striking dynamic instabilities  when put under a load~\cite{rathee2017localized, waitukaitis2012impact} -- the effects of which are seen, for example, in the rapid solidification of cornstarch suspensions that prevents a runner from sinking~\cite{van2012running}. These instabilities are only partially understood~\cite{rathee2017localized, comtet2017pairwise, saint2018uncovering}. However, it is believed that they involve rapidly propagating hydrodynamic vortex fronts that cause order-of-magnitude level viscosity fluctuations on tens of millisecond timescales~\cite{chacko2018dynamic, han2016high, comtet2017pairwise}.  Such fluctuations would be averaged out using existing measurement methods, but could be observed using our technique. Indeed, even the predicted dynamics of the most simple out-of-equilibrium system one could imagine --  a Brownian particle heated to a temperature above its environment~\cite{mazo1974theory, rings2010hot, joly2011effective} -- have not yet been experimentally verified. Measurements capable of resolving both the thermalisation of position and velocity, such as those reported here, are required to do this.

\section{Methods}

\subsection{Optical tweezers apparatus}

     Figure~\ref{fig2} illustrates our experimental setup (see Section 3 of the Supplementary Information for more details).   A monolithic Nd:YAG laser is used to produce a  low-noise 1064~nm trapping field. 
  A reference beam is split off and detected on a photodiode, with the resulting photocurrent used  in data processing to suppress laser amplitude noise in the viscosity estimate (see Supplementary Information Section 4.1).
   The trap field  is then spatially cleaned with a single mode fibre, expanded to slightly overfill the objective of a purpose built inverted microscope, and passed through the sample where it scatters off a microparticle.  After the microscope, the light is collected through a  condenser and aligned into two detection systems, a conventional split-detection system and the structured-light detection system developed to allow us to reach the ballistic regime. Split-detection is implemented using a half-mirror followed by a balanced detector.  The objective and condenser are each water immersion with 1.3 numerical aperture.

\subsection{Concept of structured-light detection}

In Fig.~\ref{fig2}{\bf a}-{\bf d} the concept of structured-light detection is introduced using the simplifying approximation that 
the transverse profile of the output field from the optical tweezer is Gaussian. The actual transverse profile has more detailed structure due to the complex scattering pattern from the microparticle. However, so long as both the input trapping field and scattering from a centred particle are symmetric on reflection about the axis of the trap, the approach remains conceptually valid:  the very weak asymmetric component of the scattered field carries all information about the particle displacement, and the much brighter symmetric component can be filtered out. In our experiments the input trapping and scattering are both symmetric to good approximation, such that the scheme is applicable. We note that, even in cases where the trapping field has some asymmetric component, this could be corrected using an appropriate offset of the waveplate from the axis of the output field, with complete rejection from single-mode fibre remaining possible in principle.

Compared with split-detection, our structured-light detection approach not only resolves the problem of power handling, but also allows the mode-shape of the detected field to be easily optimised. In split-detection, this modeshape is defined by the transverse profile of the trap field.
 Using structured detection, it can be chosen arbitrarily using holographic techniques to control the spatial component of the scattered field that enters the single mode fibre.  In our experiments, we achieve a basic version of this by controlling the magnification of the optical field prior to the fibre, as shown in Fig.~\ref{fig2}. As a concrete comparison, taking the trap field as Gaussian and approximating the information-carrying component of the scattered field as a TEM$_{01}$ transverse electromagnetic mode (as is the case if the scattering introduces a pure displacement of the trap), this approach allows a maximum detection efficiency of  $83\%$ as shown in Fig.~\ref{fig2}{\bf d}, compared to $64\%$ for split-detection~(see Section 2 of the Supplementary Information).

\subsection{Data analysis}

Care must be taken when analysing the particle trajectories to ensure that filtering does not introduce spurious velocity correlations from noise. Low pass filtering white noise, for example, introduces memory in what would otherwise be uncorrelated random noise. The data in Fig.~\ref{fig3}{\bf b}\&{\bf c} has been band-pass filtered and resampled to ensure that the correlations in {\bf c} are true velocity correlations and are not introduced from noise (see Supplementary Information Section 4.3).

As discussed earlier, as well as giving an estimate of the viscosity, the fits to the velocity power spectrum in Fig.~\ref{fig4} return an overall scaling factor. Assuming the temperature of the sample to be 23$^\circ$C, this provides the conversion factor between Volts measured on the detector and the position of the particle. Here, this fitted conversion factor is only used to express the position data in metres. It is not required in the estimation of viscosity.

When calculating the expected fluid viscosity in Fig.~\ref{fig4}, we assume the liquids to be pure and include the effects of heating from the trapping laser. We use the model from Ref.~\cite{peterman2003laser} to account for viscosity changes due to laser heating. The optical absorption $\alpha=14.2$~m$^{-1}$ and thermal conductivity $C=0.60$~W/m/K of water~\cite{peterman2003laser}, give a temperature increase from $296.1 \pm 2$~K to $297.7 \pm 2$~K. For isopropanol, $\alpha=9.1$~m$^{-1}$~\cite{sani2016spectral} and $C=0.14$~W/m/K result in a temperature of $300.5 \pm 2$~K. Optical absorption measurements were not available for methanol and acetone, for which we assume similar temperature to water.

\subsection*{Acknowledgments}
We thank Ping Koy Lam for providing the split waveplate used to implement structured detection, Nicolas Mauranyapin for taking scanning electron microscope images of the microparticles. We also acknowledge Nicolas Mauranyapin for useful discussions, along with Halina Rubinsztein-Dunlop, Isaac Lenton and Alex Terrasson. This work was supported primarily by the Air Force Office of Scientific Research (AFOSR) grant FA2386-14-1-4046. It was also supported by the Australian Research Council Centre of Excellence for Engineered Quantum Systems (EQUS, CE170100009). W.P.B. acknowledges the Australian Research Council Future Fellowship, FT140100650. M.A.T. acknowledges the Australian Research Council Discovery Early Career Research Award, DE190100641.

\newpage

\section{References} \label{Sec:References}

\bibliographystyle{plain}
\bibliographystyle{unsrt} 
\bibliographystyle{naturemag} 

\bibliography{Viscosity_refs}{}

\newpage

\part*{Supplementary Information}

\newcounter{eqnctr}
\renewcommand{\thefigure}{S\arabic{figure}}

\setcounter{section}{0}
\setcounter{figure}{0}

This Supplementary Information provides details about the theory, modelling, simulations, fitting procedures, and experimental techniques used in the paper ``\emph{Ultrafast viscosity measurement with ballistic optical tweezers}''.

\section{Theoretical modelling and simulation}
To model the experimental data we use hydrodynamic theory and simulations of the detected motion of corresponding particles including shot-noise from the light. The simulation of the particle motion is based on the equation of motion in the frequency domain \cite{berg2006power,bedeaux1974brownian},
\stepcounter{eqnctr}
\begin{equation}\tag{S\arabic{eqnctr}}
\left[-m(-i2\pi f)^2+\gamma_\mathrm{Stokes}(f)(-i2\pi f) + \kappa\right]\tilde{x}(f)=\left[2k_BT \mathrm{Re}\left(\gamma_\mathrm{Stokes}(f)\right) \right]^\frac{1}{2}\tilde{\eta}(f),
\end{equation}
where $m$ is the mass of the particle, $\kappa$ is the trap stiffness, $k_B$ is the Boltzmann constant, $T$ the temperature, Re() the real part of, $\tilde{\eta}(f)$ is the stochastic part of the thermal force in the frequency domain represented by the Fourier transform of a Gaussian random variable with mean 0 and variance 1, and the Stokes drag is \cite{berg2006power,mazur1974generalization,happel2012low}: 
\stepcounter{eqnctr}
\begin{equation}\tag{S\arabic{eqnctr}}
\gamma_\mathrm{Stokes}(f)=\gamma\left(1+(1-i)\sqrt{\frac{f}{f_f}}-i\frac{2}{9}\frac{f}{f_f}\right),
\end{equation}
with the coefficient of friction $\gamma=6\pi\eta r$, with dynamic viscosity $\eta$ and particle radius $r$. When there is a nearby surface a distance $l$ away from the particle, such as a cover-slip, the back-flow from that surface is described by Fax\'en’s correction to the friction \cite{berg2006power,mazur1974generalization}:
\stepcounter{eqnctr}
\begin{equation}\tag{S\arabic{eqnctr}}
\gamma_\mathrm{faxen}(f)=\gamma_\mathrm{Stokes}\left(1+\frac{9}{16} \frac{r}{l}\left[1-\frac{1-i}{3}\sqrt{\frac{f}{f_f}}+\frac{2i}{9}\frac{f}{f_f}-\frac{4}{3}\left(1-\exp\left(-\frac{(1-i)(2l-r)}{r\sqrt{f/f_f}}\right)\right)\right]\right).
\end{equation}
The position power spectral density (PPSD) can be obtained by substitution of the drag and re-arranging equation (S1):
\stepcounter{eqnctr}
\begin{equation}\tag{S\arabic{eqnctr}}
PPSD(f)=\frac{D}{2\pi^2f^2}\frac{1+\sqrt{f/f_f}}{(f_c/f+\sqrt{f/f_f}+f/f_i)^2+(1+\sqrt{f/f_f})^2},
\end{equation}
and the velocity spectrum,
\begin{equation*}
VPSD(f)=(2\pi f)^2PPSD(f).
\end{equation*}
\begin{table}
\begin{center}
\begin{tabular}{|c|c|c|}
\hline
Inertial Time&Fluid time&Trap time\\
\hline
$ \begin{aligned}\tau_i&=\frac{m^*}{\gamma}\\&=2.5\times10^{-7}~s\end{aligned}$&$\displaystyle\begin{aligned}\tau_f&=\frac{9m_f}{4\gamma}\\&=1.9\times10^{-7}~s\end{aligned}$&$\displaystyle\begin{aligned}\tau_t&=\frac{\gamma}{\kappa}\\&=4.5\times10^{-5}~s\end{aligned}$\\
\hline
\end{tabular}\end{center}
\caption{Characteristic times. Here $m^*=m_p+\frac{1}{2} m_f$ is the effective mass, with particle mass $m_p=1.7\times 10^{-15}$~\SI{}{\kilo\gram}  and $m_f=8.6\times 10^{-16}$~\SI{}{\kilo\gram} is the mass of the displaced fluid. $\gamma=6\pi\eta r$ is the coefficient of static friction, with dynamic viscosity $\eta=0.94$~\SI{}{\milli\pascal\second} , particle radius $r=0.59$~\SI{}{\micro\meter} and $\kappa=2.3\times 10^{-4}$~\SI{}{\newton\per\meter}  is the trap stiffness. The numerical values correspond to our experimental values.}
\end{table}

The thermal force limits the accuracy with which a part of this spectra can be determined. As there is no frequency correlation the number of independent points, $N$, within the bandwidth ($bw$) is roughly $N\approx bw t$ with $t$  being the measurement time.  As these points are uncorrelated the uncertainty in their average scales approximately as $1/\sqrt{N}$ if the source is in equilibrium. This is what sets the trend Fig. 5b (main text). 

Ideally the light that probes the particle position will have a noise floor set by the shot-noise of light. The shot-noise has an effectively infinite bandwidth. The cross-over between thermal and shot-noise sets the limit to how fast the particle can be tracked. In the results of the simulation shown in Fig. 5b (main text), the shot-noise is added to the particle position in the frequency domain with a magnitude to give the same bandwidth as obtained in the experiment and with the stochastic part represented by the Fourier transform of a random Gaussian variable.

\section{Limits on position measurement and theoretical efficiency}
The electric field at back focal plane position $X,\,Y$ after scattering from particle at a position $x,\,y,\,z$, with $y=0$, $z=z_0$, where $z_0$ is the stable trapping height is
\[
E(X,Y;x)\approx E(X,Y;0,z_0)+x\frac{\partial{E}}{\partial x}.
\]
The field can be replaced with a function of normalised spatial wavefunctions and expanded to first order:
\begin{align}\nonumber
E(X,Y;x)&=A(x)\psi(X,Y;x),\\
\stepcounter{eqnctr}\tag{S\arabic{eqnctr}}
\label{equ:mode}&\approx A(0)\psi(X,Y;0)+g x A(0) \psi'(X,Y;0),
\end{align}
where $\psi'$ is a symmetry breaking wavefunction, and $|A|^2$ is the number of photons. Using this formalism, the proportion of scattered photons that are in the information carrying component, $\psi'$, is given by $|gx|^2$. Re-arranging equation (\ref{equ:mode}) to find $|gx|^2$, we obtain:
\[
|gx|^2=\frac{\int |E(X,Y;x)-E(X,Y;0)|^2 dXdY}{\int |E(X,Y;0)|^2 dXdY}.
\]
For \SI{1064}{\nano\meter} circularly polarised light with N.A.=1.3 trapping a \SI{1.18}{\micro\meter} radius particle in water, we find for small displacements that $|g|=3.03\times10^{-4}$~\SI{}{\per\nano\meter}.

The measurement of motion into the ballistic regime requires a bandwidth of approximately \SI{5}{\mega\hertz}, and temporal resolution of \SI{100}{\nano\second}. For these parameters, the RMS displacement will be $(kT/m^* )^{1/2}t=$\SI{0.14}{\nano\meter}. For the value of $|g|$ obtained and displacement of \SI{0.14}{\nano\meter} only 1 photon in the information containing mode is generated for $6\times10^8$ unperturbed trapping photons. Measurement at \SI{100}{\nano\second} timescale therefore requires measurements with $6\times10^{15}$ unperturbed photons per second, i.e. \SI{1}{\milli\watt}, to provide only 1 photon per measurement. Since motion is unresolvable with $<1$ photon, this is a lower limit to the power which must be measured with completely optimal detection in order to detect ballistic motion. 

The detection efficiency of split detection and the structured light detection in the main Fig. 2 is calculated from the electric field overlap integral. For split detection, an approximate TEM$_{01}$ from the displaced scattered field is overlapped with the Gaussian trap field. The left and right halves of the modes are split and subtracted, so that the overlap is between the TEM$_{01}$ of the displaced scattered field and a flipped Gaussian beam. This gives an efficiency of $2/\pi=64\%$. The $83\%$ efficiency for the structured light detection is obtained by calculating the overlap between the flipped TEM$_{01}$ and the Gaussian mode of the single mode fibre as function of the waist size of the TEM$_{01}$ beam. The optimum is shown with the grey line in Fig 2 c).

\section{Experimental details and design}
\subsection{Microscope design}
For three-dimensional optical trapping of dielectric particles in aqueous solutions, we constructed an inverted microscope system designed for stability and ease of use. The light was generated from a Nd:YAG laser (1064~nm, Innolight Prometheus) and filtered and expanded as shown and described in Fig. 2 (main text). This beam is directed towards a high number objective lens (Nikon Plan Fluorite, NA 1.3, oil immersion) which is mounted on nano-stage (MAD City Labs: Nano-F200S, 0.2~nm resolution) for axial or z-axis movement. For sample placement, a 2-axis nano-stage (Nano-BioS200, precise movement) is mounted on another 2-axis micro-stage (MAD City Labs: MCL-MOTNZ, coarse movement). The whole assembly of microscope objective, axial and lateral stages are fastened on a sturdy aluminium base plate which sits on steel posts clamped to the optical table to minimise the mechanical vibrations. 

To collect the scattered light from the trapped particle, an identical objective (condenser), is mounted a threaded aluminium cylinder which provides $\sim 400$~\SI{}{\micro\meter}/rev axial displacement. This cylinder is attached to a triangular shaped plate whose design is similar to AFM stages. 

In this design, a high load carrying and high precision manual micrometer (Newport: BHC30.10, \SI{40}{\kilo\gram}, \SI{4}{\micro\meter}) is attached on each corner of the triangular plate for angular alignment control of the condenser. A steel ball-bearing connects each micrometer to the lower steel posts. At the first corner directly with grease, so it can slide in 2 dimensions. At the second corner via a linear grove that guide the bearing in 1 dimension. At the third corner, the ball-bearing head of the micrometer is placed in between three coinciding steel balls such that it is touching each ball at only one point. The steel balls are made fixed on the custom steel plate, which is mounted on two single axes high load, lockable micro-stage (Newport: M-UMR8.4) for lateral movement of the triangular plate and hence condenser. 

The sample is imaged using illumination from the top with a fibre-coupled LEDs (Thorlabs: M530F2) going through a dichroic mirror. The picture is captured below, again through a dichroic mirror with a CCD camera (AVT Manta G-031B). This whole design was made to provide 5-axis degree of freedom to the condenser to efficiently align it to the scattered light while minimising the mechanical vibrations.

\subsection{Spatial filter alignment procedure}
Following the beam path shown in Fig. 2 (main text) from the laser, a single mode polarisation maintaining fibre is used to clean the spatial optical mode. The telescope enlarges the beam to around 5\% overfilling of the objective. This value is chosen to give the highest trap stiffness and thereby particle-light interaction while minimising the distortion of the optical mode. The distance between the objectives is first chosen to collimate the output beam.

The split wave plate is placed in the collimated beam 1.5m after exciting the microscope on the optical table. This large beam minimises the optical loss from the merged half wave plate parts to around 5\%. Optimising the fibre coupling lens with the fundamental mode gives 85\% coupling into the fibre, however this position is not optimal for the signal bandwidth. From this position the fibre coupling lens is moved to 40\% coupling which empirically gives the highest bandwidth of particle thermal noise over optical noise.

\subsection{Sample production}
The sample solutions are made by adding a small amount of silica beads to a 1mL Eppendorf tube containing the fluid. We use Millipore water, Dulbecco's phosphate buffered saline, and cleaning grade isopropanol, acetone and methanol with silica particles radius \SI{1.18}{\micro\meter} and \SI{0.89}{\micro\meter} (Cospheric LLC). The particle concentration is checked in the microscope and the sample solution is typically diluted further to make it possible to find a free particle within 15 minutes, whilst also making multi-particle trapping events rare. 

The sample chambers are made from two type 0 cover slips. The edges of one cover slip is covered in a thin layer of vacuum grease. 10-\SI{20}{\micro\liter} of a sample solution is placed on the coverslip and the second coverslip is placed on top. The sample is immediately ready to be inserted in the microscope. The vacuum grease is specified to be and appears inert to all our fluids.  

\section{Data acquisition and analysis}

The power spectrum obtained from the experimental time trace is the sum the particle signal, the laser noise and the electronic detection noise. Furthermore, it is multiplied by the detection systems transfer function. For absolute viscosity measurements, the noise and the transfer function must be estimated and included in the analysis. 

The electronic noise is estimated from a calibration measurement. The laser noise floor is estimated during the main data processing loop as described below.

\subsection{Data acquisition and measurement efficiency}
The detectors in the structured-light detector detector and the reference detector, shown in Fig. 2 (main text), are amplified photo-detectors (Thorlabs pda10ec). The reference detector always has \SI{1.0}{mW} of optical power. The electronic signal of each detector is high-pass filtered with high-order \SI{100}{kHz} high-pass filters, amplified by \SI{15}{dB} using an attenuator and amplifier, low-pass filtered at \SI{60}{MHz}. The analog data is digitised on an oscilloscope (Agilent DSO-S 104A) running in high resolution mode with \SI{500}{MS/s} sampling rate. The sampling rate is chosen this high because the high resolution mode uses a boxcar filter which modifies the spectra near the sampling frequency.  

The reference and particle signal are synchronised in time in post-processing using a laser noise trace, using the cross correlation of the technical laser noise. Then the data is digitally low pass filtered and down-sampled to \SI{100}{MS/s}, to reduce the required processing power in the later data-processing steps.

A particle is trapped, typically near the lower coverslip. The particle is moved to 10-15~\SI{}{\micro\meter} above the cover slip and the fibre coupling is optimised, measured with a power meter. With an estimated 200~mW at the sample we get 100~mW at the split wave plate and 35~mW-40~mW coupled into the fibre. The split wave plate is moved to the middle of the beam where it suppresses the light. The suppression is optimised using both side to side movement and rotation around the vertical axis, typically suppressing the field to a fluctuating power around \SI{10}{\micro\watt} to \SI{30}{\micro\watt}. The split wave plate is then moved to the side to allow \SI{1}{mW} to enter the fibre, in order to perform homodyne detection of the particle signal. The fibre is then moved from the power meter to the detector and the measurements can begin.

Measurements are taken in series of 3-5 acquisitions of 0.2~s with 10-30~s delays. The suppression is checked and every 5 minutes the split wave plate is moved aside to check the coupling. After a successful set of particle trace acquisitions, the particle is released and checked that no other visible objects were trapped and the laser noise is acquired for calibration purposes.

\subsection{Transfer function}
The detector transfer function is estimated from a set of calibration measurements taken separately from the main experiment. Given the stochastic nature of the thermal force, only the amplitude response is required. Shot noise gives a perfect white noise which can be used to estimate the amplitude response. If just the shot-noise was measured, the transfer function could be directly determined as the ratio between it and the detected signal. However, at frequencies below 10 MHz there is a significant contribution from technical laser noise. The reference detector gives a copy of the technical noise, but with slightly different transfer function, optical power, electronic amplification and electronic noise. To estimate the vacuum fluctuations on the main detector underneath the technical laser noise we subtract the reference modified by the relative transfer function, power, amplification and electronic noise between the two detectors. This subtracted trace is smoothed by averaging repeated measurements and fitting a 5-th order power function polynomial. This is then inverted to give the transfer function.

\subsection{Fitting the spectrum}
The technical laser noise also reduces the bandwidth of the particle tracking. The reference detector gives a copy of the technical noise. However, at high frequencies where the shot-noise dominates the technical laser noise, the noise is necessarily uncorrelated. Subtraction of the reference detector would add the uncorrelated noise at these high frequencies, reducing the effective bandwidth. To minimise the noise contribution at all frequencies in the particle tracking, the correlation between PD2 and PD1 is calculated in 50 kHz intervals ($\Delta$f) and used to subtract the two according to :
\[
\tilde{X}_\mathrm{bal}(f)=\tilde{X}(f)-\frac{\mathrm{cov}\left(\tilde{Y}(\Delta f), \tilde{X}(\Delta f)\right)}{\mathrm{var}\left(\tilde{Y}(\Delta f)\right)}\tilde{Y}(f)
\]
with $\tilde{X}$ being the frequency domain signal from PD1 and $\tilde{Y}$ from PD2 and $f$ is within $\Delta f$.

The split detector data shown in Fig. 3 a) (main text) is re-sampled to 10 MS/s without any filtering.
To obtain time traces as shown in Fig. 3 b) and c) (main text), the data is low pass filtered with an ideal filter at 3dB bandwidth relative to the laser noise, to exclude the laser noise dominated points. The data is then down-sampled to the corresponding Nyquist sampling rate to make each point in the time domain independent. The voltage to meter conversion is performed with the factor obtained from the fitting procedure described in the main data processing loop. Still in the frequency domain, one copy of the data is converted to velocity by multiplying it by $\pm i\omega$. The data is then converted back to time domain to give the time traces shown in Fig. 3 b) and c) (main text).

For the data used in the main processing loop the position power spectrum is calculated and then multiplied by $(2\pi f)^2$ to obtain the velocity power spectrum. This is a computationally demanding process due to the large quantity of data. Both spectra are therefore averaged and resampled in 50 kHz bins which allows fast and effective processing. This is the final form of the experimental position spectrum ePPSD and velocity spectrum (eVPSD) used for fitting. The spectra plotted in Fig. 3 e) and f) have further been divided by the estimated transfer function to correct for the frequency dependence of the detection system. 

 The split detector data shown in Fig. 3 e) and f) is down-sampled to 100~MS/s and averaged in 500~Hz bins to show the low frequency part of the spectrum. The data above the bandwidth is not shown on the figure.

The 3~dB bandwidth is estimated as the minimum of the eVPSD. This value is primarily used to set an upper frequency for the fitting procedure.

For the fit, the theoretical spectrum needs to have the same processing as the experimental data. First, the shape of the spectra are calculated using the same sampling rate and measurement time as the data and using equations (S1)--(S4). The spectra are multiplied with the estimated transfer function and averaged in 50~kHz bins in the same process that was used for the experimental data. This is repeated as function of viscosity in 4000 logarithmically equal steps between 0.1 and 10~mPa~s to create a lookup table for the theoretical velocity power spectrum (tVPSD). 

The tVPSD needs a noise floor as close as possible to the noise which remains in the eVPSD to accurately compare the eVPSD and tVPSD. The shape of this noise floor is estimated from the electronic noise, the detected shape of the shot-noise, and the magnitude of the technical noise in the reference detection and the relative power in the eVPSD. 

The main fitting procedure is a fast converging iterative process between determining the magnitude of the noise floor and the viscosity. The first estimate of the magnitude of the noise floor is taken from the ePPSD at 25-30~MHz where the particle signal is small. The tVPSD is then normalised to have the same area as the eVPSD minus the estimated noise within the fitting bandwidth for all viscosities and the noise floor is added to the tVPSD. Least square fitting is then performed on this form of the tVPSD and the eVPSD to determine the best fitting viscosity $\eta_1$. Then the volt to meter conversion factor, $v2m$, is found as the ratio of tVPSD($\eta_1$) to the normalisation. Finally an updated noise floor is determined by subtracting $v2m\cdot$tPPSD($\eta_1$) from eVPSD, and the process is repeated until the viscosity converges after a few iterations. 

To make the estimation at the shorter timescales, the 0.2~s data is converted from the frequency domain to time time domain as velocity and split for independent processing. The velocity spectrum for each segment is then calculated. The noise floor is assumed to be the same as for the 0.2~s measurement. The fitting is then performed directly.

\subsection{Uncertainty in the measurements}

\begin{figure}
\begin{center}
\includegraphics[width=0.8\textwidth]{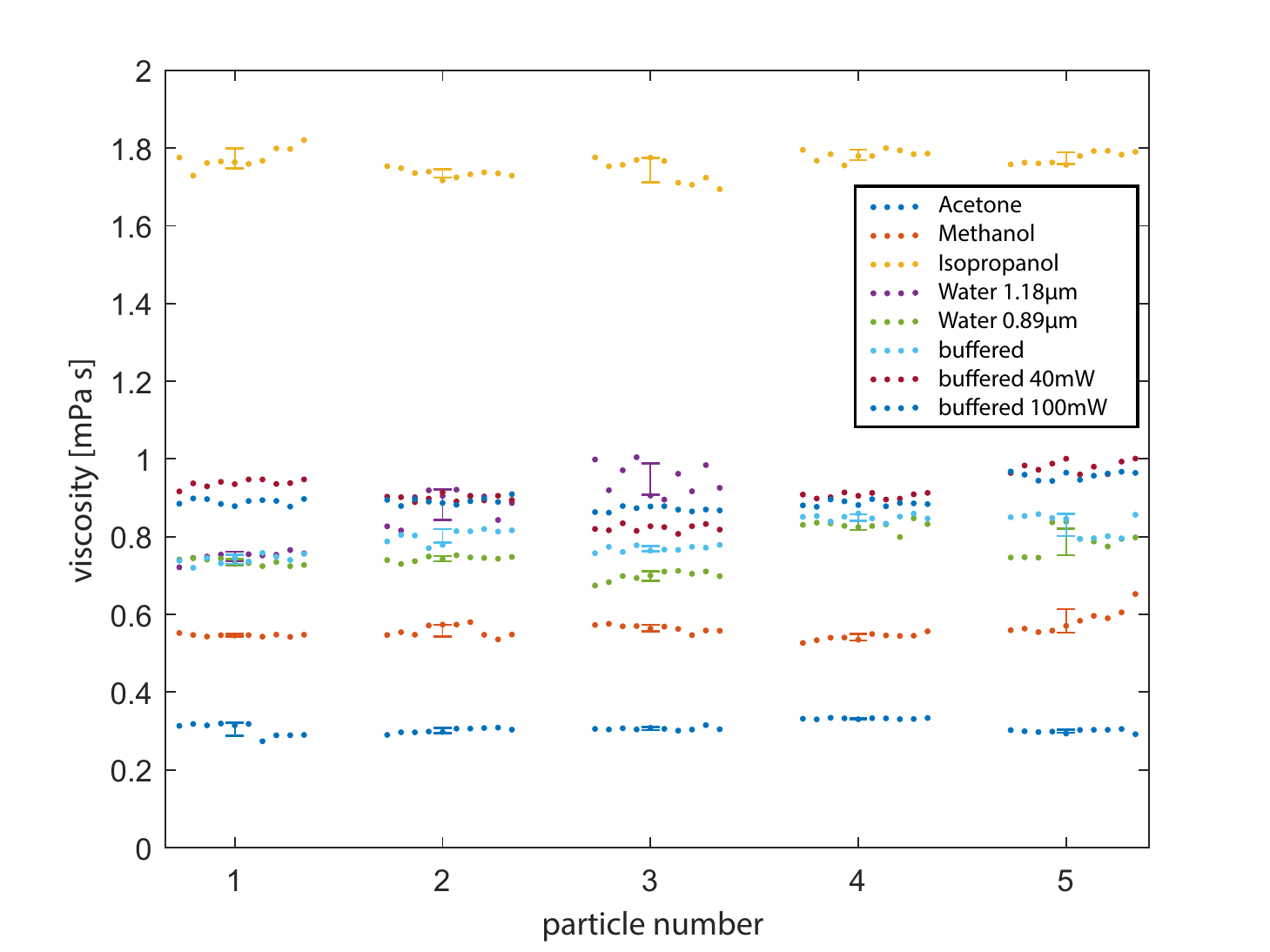}
\end{center}
\caption{\label{fig:visco} Individual viscosity measurements that are summarized in Fig 4 b) of the main text. Further data is shown taken at lower trapping power in the buffered solution. For each medium 5 particles were trapped and 10 measurements taken per particle, with the exception of water \SI{1.18}{\micro\meter} where only 3 particles were used.
The errorbars indicate the uncertainty for each particle.
}
\end{figure}

Results of the fitting procedure are shown in Fig. \ref{fig:visco} for several different trapped particles and solutions, and for multiple measurements on the same particle as indicated in the legend. The  viscosity is estimated over 0.2~s. 
 Viscosity measurements are consistent in each time series and for each particle, we obtain the viscosity of acetone ($\eta\simeq0.30\times10^{-3}$\SI{}{\milli\pascal\second}), methanol ($\eta\simeq0.63\times10^{-3}$\SI{}{\milli\pascal\second}), water ($\eta\simeq0.85\times10^{-3}$\SI{}{\milli\pascal\second}) and isopropanol ($\eta\simeq1.76\times10^{-3}$\SI{}{\milli\pascal\second}). Despite the good degree of consistency, small variations are present. We attribute these fluctuations to two main sources in roughly equal measure: Particle size and geometry variation within the samples provided by the manufacturer, and the presence of small particulates which are commonly found in all but the solutions of highest purity. 
 We found that most of the samples used in the experiments contained small particulates due to the sample preparation in a non-clean room environment. These particulates were far-sub-wavelength in size and were attracted by the optical trap. This interfered with the measurement of the probe, causing the observed viscosity fluctuations.
 The statistical uncertainty arising from thermal- and shot-noise  is significantly smaller than the contributions from particle variability and particulates. For most fluids measured the variation of viscosity was no more than \SI{0.1}{\milli\pascal\second}.

The effect of variation of the particle size and geometry on viscosity can be seen in Fig. \ref{fig:visco}, by comparing data points that keep the fluid medium and the manufacturers nominal particle size fixed, but change the trapped particle. The effect of small particulates can be seen by comparing data points 
 in the same scenario but using multiple measurements on the same identical trapped particle.

\bibliographystyle{unsrt}

\end{document}